# UniToBrain dataset: a Brain Perfusion Dataset

Daniele Perlo[1][0000−0001−6879−8475], Enzo Tartaglione[2][0000−0003−4274−8298], Umberto Gava[3][0000−0002−9923−9702], Federico D'Agata[3], Edwin Benninck[4], and Mauro Bergui[3][0000−0002−5336−695X]

[1] Fondazione Ricerca Molinette Onlus
[2] LTCI, Télécom Paris, Institut Polytechnique de Paris
[3] Neuroscience Department, University of Turin
[4] University Medical Center Utrecht

**Abstract.** The CT perfusion (CTP) is a medical exam for measuring the passage of a bolus of contrast solution through the brain on a pixel-by-pixel basis. The objective is to draw "perfusion maps" (namely cerebral blood volume, cerebral blood flow and time to peak) very rapidly for ischemic lesions, and to be able to distinguish between core and penumbra regions. A precise and quick diagnosis, in a context of ischemic stroke, can determine the fate of the brain tissues and guide the intervention and treatment in emergency conditions.

In this work we present UniToBrain dataset, the very first open-source dataset for CTP. It comprises a cohort of more than a hundred of patients, and it is accompanied by patients metadata and ground truth maps obtained with state-of-the-art algorithms. We also propose a novel neural networks-based algorithm, using the European library *ECVL* and *EDDL* for the image processing and developing deep learning models respectively. The results obtained by the neural network models match the ground truth and open the road towards potential sub-sampling of the required number of CT maps, which impose heavy radiation doses to the patients.

## 1 Introduction and Clinical Background

The occlusion of a cerebral vessel causes a sudden decrease in blood flow in the surrounding vascular territory, in comparison to its centre. The identification of such an occlusion reliably, quickly and accurately is crucial in many emergency scenarios like ischemic strokes [7]. The CT perfusion (CTP) is a medical exam for measuring the passage of a bolus of contrast solution through the brain on a pixel-by-pixel basis. CT perfusion is performed with a sampling time of approximately 1 Hz. Based on these measurements, low-dose serial scans are acquired from which blood time-density curves and several other parametric maps are calculated. The blood time-density curves correspond to the passage of the contrast agent into the brain tissue. The most relevant parameters used in clinical practice are maps representing Cerebral Blood Volume, Cerebral Blood Flow and Time To Peak (called CBF, CBV, TTP respectively) [1].

Ischemic lesions develop very rapidly, originating from the central area of the occluded vascular region and progressively expanding to increasingly peripheral regions. From the onset of symptoms, two different regions can be identified in the problematic ischemic area of the brain. The area of irreversible damage is the central region called *core*. The visible area of potential recovery and possible re-canalization, provided by the occluded vessel, is the peripheral *penumbra* region. Generally, CBV maps are used for core segmentation, while CBF and TTP for penumbra areas. Therefore, the identification of the core and of the penumbra can predict the fate of the brain tissue itself and guide physicians in re-perfusion treatments [7, 17]. An example of colored perfusion maps and core-penumbra region is shown in Fig 1. The extent of the core and its penumbra area can be estimated clinically based on symptoms and their time of onset, and using common perfusion techniques such as CTP.

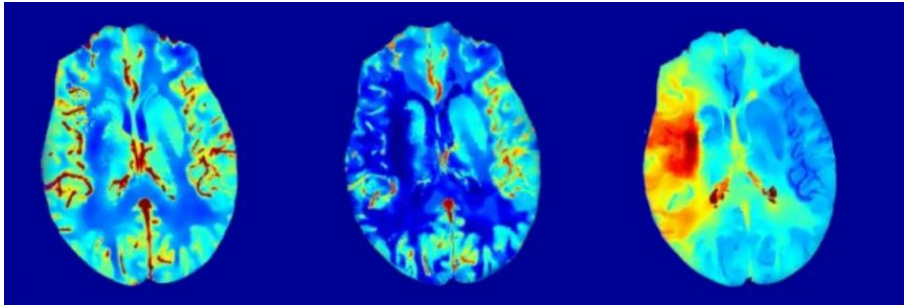

Fig. 1: Examples of perfusion maps, from left to right CBV, CBF and TTP. While core is the dark blue region in CBV, the penumbra is evident in CBF and TTP: in TTP it is particularly evident, in red.

A time-intensity curve is the result of processing the signal intensity values over time. Several algorithms are used to perform deconvolution of time-intensity curves, but some of them are not public and may produce maps with divergences [12]. If we place ourselves in an ideal environment, where we find limited noise, limited variance and no artifacts created by patient motion during scanning, the optimal choice for obtaining realistic, accessible and reproducible maps is an analysis performed pixel by pixel, such as the one performed by deconvolution-based algorithms. Unfortunately, in the real clinical case, the information must be redundant in order to overcome the problems caused by the presence of noise, large variances and motion artifacts. In practice, this results in obtaining more brain slices, requiring more patient X-ray exposition, more acquisitions, estimation of an Arterial Input Function (AIF), and a series of spatial pre-processing steps for noise and variance reduction.

## 2   DeepHealth Related Works

The European Union is spending significant amount of resources properly leading the AI development and research to maintain its proper position in this emergent and challenging context. EU funds projects to deploy pilots and large-scale experimental research solutions by combining the latest discovered technologies in artificial intelligence. The same effort is spent to support AI technologies like distributed high-performance computing, cloud computing and big data support. *DeepHealth* [6, 13] is an example of this effort, that is thought (and developed) as a health-focused project. CTP data gathered for UniToBrain dataset are part of a larger collection of open-access datasets [3, 14] developed for the DeepHealth project. The European framework of *DeepHealth*, called *DeepHealth Toolkit*, is a single entity that aims to use heterogeneous architectures, such as high-performance computing, big data and cloud computing, to provide deep learning capabilities and computer vision off-the-shelf services. *DeepHealth Toolkit* 's purpose is to make it easier to develop and deploy new applications that solve specific problems regardless of the type of context or application field. Our contribution is achieved with the support the *DeepHealth Toolkit*, by taking into account some of its main components, in particular *EDDL* and *ECVL* libraries.

### 2.1   EDDL

*EDDL* is a deep learning library that was originally developed to be general-purpose and it meets the AI development needs inside the DeepHealth context, therefore with the aim to be used in healthcare applications. *EDDL* is a free and open source software library, available on online repository as a core part of the *DeepHealth Toolkit*. It integrates and supports the most popular and well known Artificial Neural Network (ANN) topologies, by including both convolutional and recurrent models. *EDDL* aims to simplify the development of hardware-specific accelerated deep learning mechanisms by providing hardware-independent tensor operations and ANN topologies components like activation functions, regularization functions, optimization methods, and multiple types of different layers. The neural network library contains both high-level tools, such as training and evaluation of a model, and low-level tools that allow finer control of each step in the training or inference loop, like individual epochs, batches, or gradients manipulation.

### 2.2   ECVL

The *European Computer Vision Library (ECVL)* is the *DeepHealth* computer vision library, developed to be general-purpose, with a special focus on supporting healthcare applications. It provides high-level computer vision capabilities, by providing also specialized and hardware accelerated implementations of commonly used image processing algorithms in deep learning. *ECVL*'s design revolves around the pivotal concept of Image as the centerpiece of the entire library.

Special features are included in the library for medical image data manipulation. These images are often retrieved with proprietary or multi-scan formats, such as DICOM, NIfTI and others virtual slide formats. Image provides the arithmetic instruments in order to apply mathematical operations between images and scalars. More common affine image transformations, such as rotating, scaling, mirroring, and color space changes are available. *ECVL* is accompanied by a *Python API* called pyecvl, just like pyeddl, whose main advantage is that it not only speeds up application developments but also integrates with the rich *Python* ecosystem of scientific programming tools. Support for *NumPy* allows developers to process data with many other scientific tools. pyecvl, as well as other toolkits, is available as open-source software.

## 3 The *UniToBrain* Dataset

The University of Turin (UniTo) released the open-access dataset UniToBrain collected within DeepHealth project [6]. UniToBrain [9] is a dataset of Computed Tomography (CT) perfusion images (CTP). The dataset includes 258 patients from multiple health institutions. Perfusion data were obtained from hospital Picture archiving and communication system (PACS) of *Città della Salute e della Scienza di Torino* by Neuroradiology Division in Molinette Hospital, by doing a retrospective research.

All the data collected within this study are retrieved with procedures complying with institutional ethical standards, the 1964 Declaration of Helsinki and its subsequent amendments, or comparable ethical standards. In addition, the requirement for written informed consent was waived because this was a retrospective study.

Perfusion maps, including CBF, CBV, TTP, were calculated using a standard spatial pre-processing pipeline followed by a nonlinear regression (NLR) method based on a state-of-the-art fast model developed and described by Bennink *et al.* [4]. A motion correction is required and was performed using a rigid registration method. Next, all images were pre-processed by implementing a bilateral filter [11]. Estimates of the AIF and vein output function (VOF) were made automatically using a sample of 100 voxels. The impulse response function (IRF) of perfused tissue is described by the model developed by Bennink *et al.*, in terms of CBV and tracer delay, which is fundamental in the clinical context of ischemic stroke. The tissue temporal attenuation curve and associated maps (CBV, CBF, and TTP) are then estimated using the AIF and IRF calculated by that method [4]. Along with the dataset, we provide some utility files, by using the *DeepHealth Toolkit*, in order to pre-process the DICOM files, the ground-truth maps and to load the dataset for ANN training purposes.

### 3.1 *DeepHealth Toolkit* integration

The *DeepHealth Toolkit* is a versatile deep learning system with fully integrated image processing and computer vision capabilities that uses HPC and cloud in-frastructure to perform parallel and decentralized AI inference and learning processes. The toolkit is designed to handle large and growing datasets, by matching the nature of medical image datasets. To that end, the *DeepHealth Toolkit* solution aims to transparently integrate the latest parallel programming technologies to leverage the parallel performance capabilities of HPC and cloud infrastructures, by including symmetric multiprocessors (SMPs), graphics processing units (GPUs) and various other computing acceleration technologies. The toolkit also includes features that can be used for training and inference by abstracting the complexity of different computational resources and facilitating architecture developments adopted in the learning and inference phases. The following scripts for data management are then provided along with the dataset:

dicomtonpy.py: It converts the DICOM files in the dataset to *NumPy* arrays. These are 3D arrays, where CT slices at the same height are piled-up over the temporal acquisition. Here pyecvl is used in order to convert and resize the ground-truth images.

dataloader_pyeddl.py: Dataloader for the pyeddl deep learning framework. It converts the numpy arrays in normalized tensors, which can be provided as input to standard deep learning models using the European library *EDDL*. The reader can look at https://github.com/EIDOSlab/UC3-UNITOBrain to have a full companion code where a U-Net model is trained over the dataset.

dataloader_pytorch.py: Dataloader for the *PyTorch* deep learning framework. The behaviour is the same of dataloader_pyeddl.py.

## 4   Methodology

Employing Deep Learning approaches to the problem of deconvolving time-intensity curves, offers several potential advantages over canonical algorithms. Indeed, DL allows the extraction of information and features that are relatively insensitive to noise, misalignment and variance. Gava *et al.*[10] shows whether a properly trained Convolutional Neural Network (CNN), based on a U-Net-like structure, on a pre-processed dataset of CTP images, can generate clinically relevant parametric maps of CBV, CBF and time to peak TTP.

Since this specific model has been originally thought for medical images segmentation [15], and in this case the goal is to generate parametric images, some changes to the standard model are introduced. There is evidence that this CNN model is effective in other tasks, other than image segmentation, as well [8]. In contrast to similar state-of-the-art U-Net based methods, no extra information, like the above mentioned arterial input function AIF, has been provided to the U-Net model.

### 4.1   Data Preprocessing

The raw pre-processing inputs are a collection of multiple CT scans and multiple targets maps for each patient, in *DICOM* format. *DICOM* is an acronym for *Digital Imaging and Communications in Medicine* and it is the standard for the communication and management of medical imaging information and related metadata. Each target map corresponds to the output of the perfusion scan at a certain scan height of the patient's brain. In order to synthesize a representative map of the patient's blood flow over time, CT scans are performed at different time points. Thus, given a target perfusion, we have to search between the patient CT scans for all the images acquired at the same height. All the CT scans for the same patient are ordered following their acquisition time point. Therefore CT images can be viewed as 3D tensors, with the third dimension being the time axis. In practice, the input pre-processing consists in finding all grayscale images correlated by the scan height and grouping them in a single image tensor, whose depth has to be the number of available CT scan for a given target map. All pixel values in the image tensor drop in range [0; 1]. Fig. 2 summarizes this input pre-processing step.

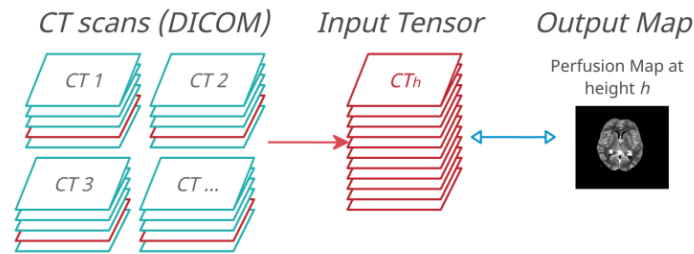

Fig. 2: A representation of the *UniToBrain* input pre-processing step. CT scans at the same scan height $h$ are staked together in order to create U-Net input tensor.

### 4.2 Network Architecture

The pre-processed images are the only input for the trained U-Net network architecture [15]. Here, we consider using it for inference on perfusion maps. The overall model can be found in Fig. 3. This particular neural network architecture has been originally developed for image segmentation: however, it proved to be effective also to solve other image generation tasks. For our purpose, we reproduce the suggested modifications [10] at the architectural level to fit the problem. For segmentation tasks, a common state-of-the-art choice is to use max-pooling layers for sub-sampling. This operator, however, introduces a non-linear behavior which prevents the forward propagation of great part of the information content [16]. According to the paper, we have used average pooling layers in place of max pool since sparse features are not expected to be extracted. A sigmoid function is used to estimate each pixel intensity for the CNN output image. Hence, depending on the chosen time granularity, the number of input channels changes accordingly.

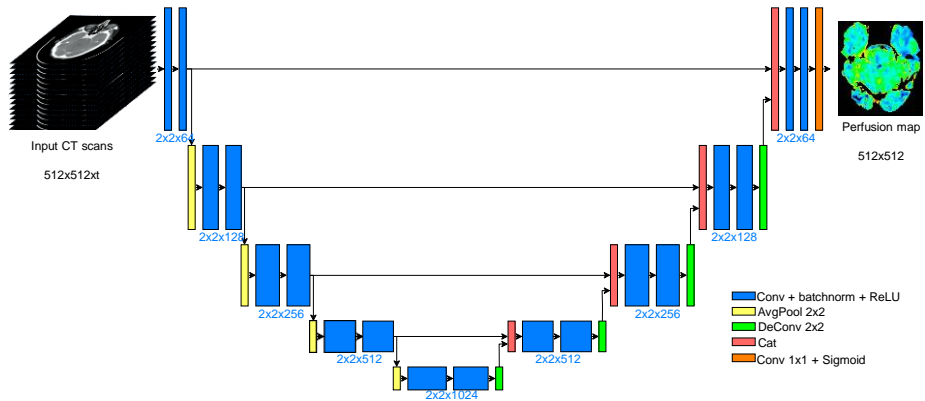

Fig. 3: A representation of the U-Net architecture used to generate perfusion maps. The model takes as input the scans acquired in different time instants. The images resolution is 512×512. After four encoding stages, the decoder stages produce a map with the same input resolution.

### 4.3  Training Procedure

The tensors, that are produced in the pre-processing step, are the only input provided to the trained U-Net architecture. In order to train our CNN, we use a widely used loss function definition for image generation tasks. We are using U-Net for perfusion map inference: instead of using standard cross entropy loss, or dice score/focal loss, which are typical for training segmentation tasks, we minimize the Mean Squared Error (MSE) loss between the required ground-truth and the CNN output. Additional information or data (e.g. AIF) are not provided to the CNN: all information is found in the pre-processed input CT images. The model is pre-trained to produce a 128×128 output map for 250 epochs, after that it is trained and tuned for 50 epochs to produce 512 ×512 maps, in which all pixel values fall into the range [0; 1]. The choice to pre-train the U-Net model with down-sampled images comes mainly from practical factors. Training on full resolution images requires a high consumption of GPU memory and the time needed to reach model convergence becomes very high. Training on down-sampled images allows to reach a solution in a much shorter time, since the task is simplified. The pre-training step produces a suitable set of starting parameters for the model: this type of approach is often used for training generative networks, and the results are comparable, and smoother, than those produced with very high dimensional image training. The training of the entire model for each target is done using the Adam optimization strategy, with a learning rate of $10^{-5}$ and batch size equals to 8.

## 5  Results and Discussion

The whole UniToBrain collection is composed from patients that differ not only in the number of exams or in the head positioning, but also their CT scans are acquired at different scan heights or scan angle. A sub-sample of 100 training subjects, and 15 for testing, was used in a submitted publication for the training and the testing of a Convolutional Neural Network [10]. This sub-sample is composed by patients from Molinette hospital only, each of them share the same features in terms of number of CT scans and acquisition parameters. The CTP acquisitions were performed using a GE 64 Scanner, and the parameters for the examination were defined as follows, the same for each case: 80 kV, 150 mAs, 44.5 sec duration, 89 volumes (40 mm axial coverage), injection of iodinated contrast agent for 40 ml (300 mg/ml) at 4 ml/s speed. In this sub-sample, 8 perfusion maps are available for each target (CBV, CBF, and TTP) for each patient. Furthermore, each sample share the same number of CT scans: the same brain portion appears 89 times at different scan timing. Therefore, the number of scans that are stacked together to compose the U-Net input is the same. In our experimentation, we use the same data sub-sample.

We evaluate the concurrent pyeddl implementation on the UniTo HPC environment (HPC4AI [2]) with four 16 GB GPUs NVIDIA Tesla T4. All simulations with *EDDL* were performed using the "low memory" option available on the neural model building phase. Due to the size of the inputs, $512 \times 512 \times 89$ wide tensors, 2 samples are computed for each GPU for each training step.

|     | Pearson index | DICE coefficient (binarized masks) |
|-----|---------------|-----------------------------------|
| CBV | 0.73          | 0.75                              |
| CBF | 0.83          | 0.52                              |
| TTP | 0.87          | 0.54                              |

Table 1: Correlation measures between full resolution U-Net outputs and related targets.

We propose the results for the three identified target perfusions. We report both MSE graphs from the pre-training phase at $128 \times 128$ resolution and from the fine-tuning step at $512 \times 512$ resolution, for 250 and 50 training epochs respectively. The evaluation of the resulting images is performed by the value of the loss Mean Square Error (MSE) function. MSE clearly indicates how much the predicted maps deviate from the target. Fig. 4b reports that MSE reach promising values below 0.01. While CBF and CBV experiments show similarities in loss slopes in both phases, predicted TTP map has an higher mean MSE over the training loop. MSE shows a slighter decrease for CBF and CBV curves, that suggests a lower gain for the pre-training at a lower resolution $128 \times 128$ pixels. The evaluation of each generated perfusion map, whose pixel values fall in the range [0 : 1], is proposed by measuring the correlation between the CNN output and the ground-truth with the Pearson correlation index. We also evaluate the highlighted brain regions consistency for each map by calculating the Dice coefficient. In order to do that, we simply binarize the perfusion maps by threshold at half intensity range. We summarize the measurements in Tab. 1.

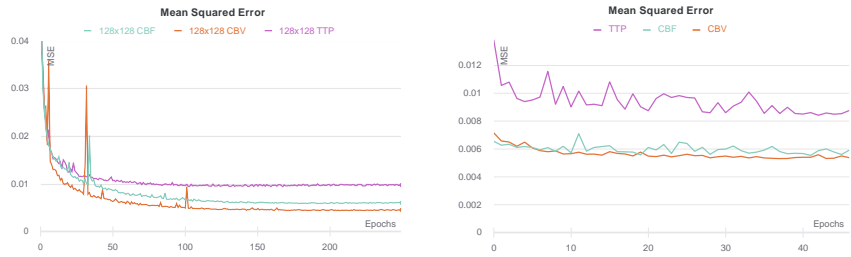

(a) Measurements at pre-training phase, for 128×128 pixels images.

(b) Measurements at fine-tuning phase, for 512×512 pixels images.

Fig. 4: MSE loss metric during training: low is better, it indicates that the generated maps are numerically very close to the targets.

Dice score suggests that CBF and TTP predictions tend to slightly over-estimate high-intensity regions. In particular, comparing the results with the Pearson index, we observe that an hard thresholding to half the intensity is not a good approach for the complex temporal and flow-related perfusion maps: for this reason, direct optimization of binarized images looks not to be a promising approach for this specific task. On the contrary, high Pearson indices suggest the success of the MSE optimization. Visual examples on the generated perfusion maps are reported in Fig. 5. In according to a team of 3 Molinette clinicians, the U-Net CNN approximates sufficient perfusion mismatch in brain tissues to create an information-rich perfusion map for patients with ischemic lesions. Clinicians analysis refer a good agreement between the CNN proposed method and the ground-truth in estimating hypo-perfused regions on CTP maps.

## 6 Conclusion

The growing potential of machine learning-based perfusion analysis methods have the potentialities to lead to new improved acquisition protocols and reduced radiation doses. In this work we introduce UniToBrain, an open-source dataset for CTP, which allows us to evaluate a state-of-the-art DL approach for generate brain blood flow maps. We have observed that CNN models can accurately evaluate the perfusion parameters by using CT scans images only, without any other input information, while the standard decomposition-based methods for CT and MRI require AIF curves measured in the saphenous vein [5]. This suggests that CNNs are capable of combining the information associated with both tissue and arterial signals to estimate CBV, CBF, and TTP maps. In this work we propose as well a baseline CNN model, derived from U-Net, implemented and open-sourced using the new *DeepHealth Toolkit*.

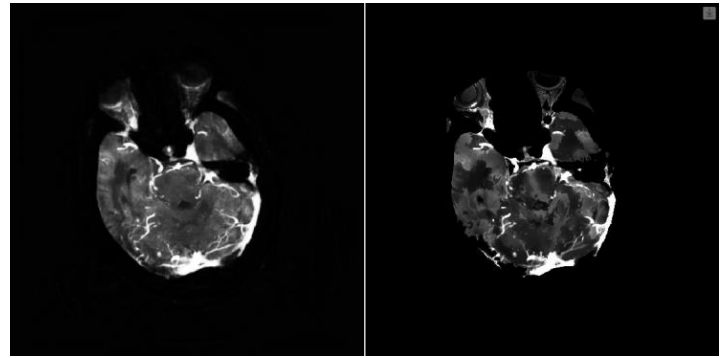

(a) CBF

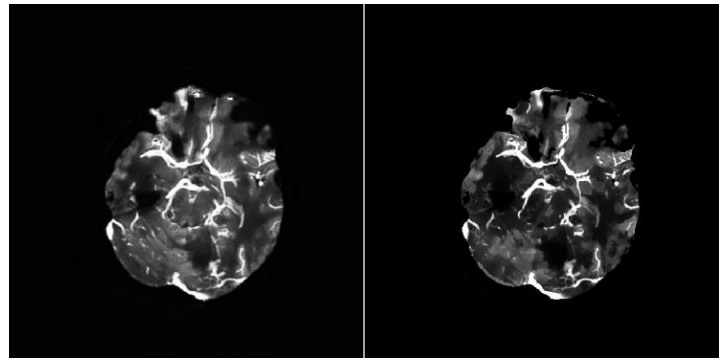

(b) CBV

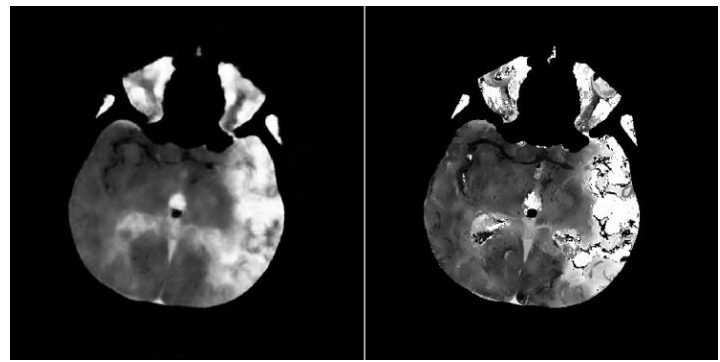

(c) TTP

Fig. 5: Examples of map predictions for each target type. The image on the left represents the U-Net output, while the ground-truth map is on the right.

## Acknowledgement

This work has received funding from the European Union's Horizon 2020 research and innovation programme under grant agreement No 825111, *DeepHealth* Project.

## References

1. Albers, G.W., Marks, M.P., Kemp, S., Christensen, S., Tsai, J.P., Ortega-Gutierrez, S., McTaggart, R.A., Torbey, M.T., Kim-Tenser, M., Leslie-Mazwi, T., Sarraj, A., Kasner, S.E., Ansari, S.A., Yeatts, S.D., Hamilton, S., Mlynash, M., Heit, J.J., Zaharchuk, G., Kim, S., Carrozzella, J., Palesch, Y.Y., Demchuk, A.M., Bammer, R., Lavori, P.W., Broderick, J.P., Lansberg, M.G.: Thrombectomy for stroke at 6 to 16 hours with selection by perfusion imaging. New England Journal of Medicine **378**(8), 708–718 (Feb 2018). https://doi.org/10.1056/nejmoa1713973
2. Aldinucci, M., Rabellino, S., Pironti, M., Spiga, F., Viviani, P., Drocco, M., Guerzoni, M., Boella, G., Mellia, M., Margara, P., Drago, I., Marturano, R., Marchetto, G., Piccolo, E., Bagnasco, S., Lusso, S., Vallero, S., Attardi, G., Barchiesi, A., Colla, A., Galeazzi, F.: Hpc4ai: An ai-on-demand federated platform endeavour. In: Proceedings of the 15th ACM International Conference on Computing Frontiers. p. 279–286. CF '18, Association for Computing Machinery, New York, NY, USA (2018). https://doi.org/10.1145/3203217.3205340
3. Barbano, C.A., et al.: Unitopatho, a labeled histopathological dataset for colorectal polyps classification and adenoma dysplasia grading. In: 2021 IEEE International Conference on Image Processing (ICIP). pp. 76–80 (2021). https://doi.org/10.1109/ICIP42928.2021.9506198
4. Bennink, E., Oosterbroek, J., Kudo, K., Viergever, M.A., Velthuis, B.K., de Jong, H.W.A.M.: Fast nonlinear regression method for CT brain perfusion analysis. Journal of Medical Imaging **3**(2), 026003 (Jun 2016). https://doi.org/10.1117/1.jmi.3.2.026003
5. Campbell, B.C., Yassi, N., Ma, H., Sharma, G., Salinas, S., Churilov, L., Meretoja, A., Parsons, M.W., Desmond, P.M., Lansberg, M.G., Donnan, G.A., Davis, S.M.: Imaging selection in ischemic stroke: Feasibility of automated CT-perfusion analysis. International Journal of Stroke **10**(1), 51–54 (Oct 2014). https://doi.org/10.1111/ijs.12381
6. DeepHealth: Deep-learning and hpc to boost biomedical applications for health (2019), https://deephealth-project.eu/
7. Donahue, J., Wintermark, M.: Perfusion CT and acute stroke imaging: Foundations, applications, and literature review. Journal of Neuroradiology **42**(1), 21–29 (Feb 2015). https://doi.org/10.1016/j.neurad.2014.11.003
8. Falk, T., Mai, D., Bensch, R., Özgün Çiçek, Abdulkadir, A., Marrakchi, Y., Bohm, A., Deubner, J., Jäckel, Z., Seiwald, K., Dovzhenko, A., Tietz, O., Bosco, C.D., Walsh, S., Saltukoglu, D., Tay, T.L., Prinz, M., Palme, K., Simons, M., Diester, I., Brox, T., Ronneberger, O.: U-net: deep learning for cell counting, detection, and morphometry. Nature Methods **16**(1), 67–70 (Dec 2018). https://doi.org/10.1038/s41592-018-0261-2
9. Gava, U., et al.: UniToBrain (2022). https://doi.org/10.21227/x8ea-vh16
10. Gava, U.A., D'Agata, F., Tartaglione, E., Grangetto, M., Bertolino, F., Santonocito, A., Bennink, E., Bergui, M.: Neural network-derived perfusion maps: a model-free approach to computed tomography perfusion in patients with acute ischemic stroke (2021). https://doi.org/https://doi.org/10.1101/2021.01.13.21249757
11. Klein, S., Staring, M., Murphy, K., Viergever, M., Pluim, J.: elastix: A toolbox for intensity-based medical image registration. IEEE Transactions on Medical Imaging **29**(1), 196–205 (Jan 2010). https://doi.org/10.1109/tmi.2009.2035616
12. Kudo, K., Sasaki, M., Yamada, K., Momoshima, S., Utsunomiya, H., Shirato, H., Ogasawara, K.: Differences in CT perfusion maps generated by different commercial software: Quantitative analysis by using identical source data of acute stroke patients. Radiology **254**(1), 200–209 (Jan 2010). https://doi.org/10.1148/radiol.254082000
13. Oniga, D., et al.: Applications of AI and HPC in health domain. In: HPC, Big Data, AI Convergence Toward Exascale: Challenge and Vision, chap. 11. CRC Press, Taylor & Francis Group (2021), ISBN: 9781032009841
14. Perlo, D., Renzulli, R., Santinelli, F., Tibaldi, S., Cristiano, C., Grosso, M., Limerutti, G., Grangetto, M., Fonio, P.: UniToChest (2022). https://doi.org/10.5281/zenodo.5797912
15. Ronneberger, O., Fischer, P., Brox, T.: U-net: Convolutional networks for biomedical image segmentation. In: Lecture Notes in Computer Science, pp. 234–241. Springer International Publishing (2015). https://doi.org/10.1007/978-3-319-24574-4_28
16. Sabour, S., Frosst, N., Hinton, G.E.: Dynamic routing between capsules. In: Proceedings of the 31st International Conference on Neural Information Processing Systems. p. 3859–3869. NIPS'17, Curran Associates Inc., Red Hook, NY, USA (2017). https://doi.org/10.5555/3294996.3295142
17. Wannamaker, R., Guinand, T., Menon, B.K., Demchuk, A., Goyal, M., Frei, D., Bharatha, A., Jovin, T.G., Shankar, J., Krings, T., Baxter, B., Holmstedt, C., Swartz, R., Dowlatshahi, D., Chan, R., Tampieri, D., Choe, H., Burns, P., Gentile, N., Rempel, J., Shuaib, A., Buck, B., Bivard, A., Hill, M., Butcher, K.: Computed tomographic perfusion predicts poor outcomes in a randomized trial of endovascular therapy. Stroke **49**(6), 1426–1433 (Jun 2018). https://doi.org/10.1161/strokeaha.117.019806